\documentclass{article}

\usepackage{amsfonts,amsmath,amsthm,amssymb}
\usepackage{mathrsfs}
\usepackage{microtype}
\usepackage{graphicx}
\usepackage{subcaption}
\usepackage{booktabs} % for professional tables
\usepackage{tikz}

\usetikzlibrary{shapes.geometric}
\usetikzlibrary{shapes.misc}
\usetikzlibrary{decorations.markings}
\usetikzlibrary{decorations.pathmorphing}
\usetikzlibrary{arrows}
\usetikzlibrary{shapes.callouts}
\usetikzlibrary{decorations.shapes}
\usetikzlibrary{fadings,decorations.pathreplacing}
\usetikzlibrary{positioning}
\tikzset{cross/.style={cross out, draw=black, fill=none, minimum size=2*(#1-\pgflinewidth), inner sep=0pt, outer sep=0pt}, cross/.default={2pt}}

\newcommand{\id}{\mathbf{1}}
\DeclareMathOperator{\Det}{Det}
\newcommand{\diag}{\mathrm{diag}}
\newcommand{\Sp}{\text{Sp}}
\newcommand{\NN}{\text{NN}}

\usepackage{hyperref}

\usepackage[accepted]{icml2019_phys4dl}

\usepackage[noabbrev]{cleveref}
\icmltitlerunning{Learning Symmetries of Classical Integrable Systems}

\begin{document}

\twocolumn[
\icmltitle{Learning Symmetries of Classical Integrable Systems}




\begin{icmlauthorlist}
\icmlauthor{Roberto Bondesan}{qti}
\icmlauthor{Austen Lamacraft}{cam}
\end{icmlauthorlist}

\icmlaffiliation{qti}{Qualcomm AI Research, Amsterdam, NL.}
\icmlaffiliation{cam}{TCM Group, Cavendish Laboratory, University of Cambridge, J. J. Thomson Ave., Cambridge CB3 0HE, UK}

\icmlcorrespondingauthor{Roberto Bondesan}{r.bondesan@gmail.com}

\icmlkeywords{Machine Learning, ICML}

\vskip 0.3in
]



\printAffiliationsAndNotice{}  % leave blank if no need to mention equal contribution

\begin{abstract}

%This document provides a basic paper template and submission guidelines.
%Abstracts must be a single paragraph, ideally between 4--6 sentences long.
%Gross violations will trigger corrections at the camera-ready phase.

The solution of problems in physics is often facilitated by a change of variables. In this work we present neural transformations to learn symmetries of Hamiltonian mechanical systems. Maintaining the Hamiltonian structure requires novel network architectures that parametrize symplectic transformations. We demonstrate the utility of these architectures by learning the structure of integrable models. Our work exemplifies the adaptation of neural transformations to a family constrained by more than the condition of invertibility, which we expect to be a common feature of applications of these methods.

\end{abstract}

% 1 page (2 cols)
\section{Problem statement}

Symmetries play a paramount role in nature and are foundational
aspects of both theoretical physics and deep learning. Knowing
symmetries of a physical problem is often the first step towards its
solution.  In machine learning, models that exploit symmetries of
their data domain are most successful at their tasks, as exemplified
by the success of convolutional neural networks for a variety of
problems \cite{DLBook}. Recently, many deep learning papers have
explored the concept of symmetry using tools from theoretical physics,
e.g.~\cite{mallat2016,Bronstein:2017aa,cohen2018, higgins2018}.

In most cases, these works assumed that the symmetries of the problem
were manifest and built neural architecture that respected those
symmetries, in order to learn more efficiently. In some situations,
however, the symmetries of the problem are hidden from us, and much of
the work is done to uncover those symmetries. A famous example is Kepler's inference from astronomical data that planetary orbits form ellipses with the sun at one focus. The fact that orbits in an inverse square law of force generically close is a consequence of a subtle symmetry of the problem that gives rise to the conserved Laplace--Runge--Lenz vector \cite{goldstein}.

Here, we present a data--driven way to learn such symmetries. %To the best of our knowledge, this is the first time that such a program is carried out.
While we concentrate here on models of Hamiltonian mechanics, we hope that the tools we develop will inspire research into symmetry learning more generally.

Learning a symmetry means learning a transformation from the original physical variables to a new set of variables in which the symmetry is manifest. Neural models describing bijective mappings are the subject of recent work on normalizing flows \cite{Rezende:2015aa,NICE,realNVP} and RevNets \cite{Gomez:2017aa,Jacobsen:2018aa}. Normalizing flows are usually constructed to have tractable Jacobians. In Hamiltonian mechanics \emph{symplectic} (or \emph{canonical}) transformations have a special role \cite{Saletan:1998aa}. Such transformations are volume preserving but have further restrictions (see \cref{eq:symp_cond} below), and so require new network architectures. Hamiltonian time evolution is itself a symplectic transformation, so these methods may be applied to neural variational inference with Hamiltonian Monte Carlo (HMC) \cite{neal}, discussed in several recent papers \cite{Salimans:2014aa,Wolf:2016aa,Levy:2017aa,Caterini:2018aa,Hoffman:2019aa}. \textbf{Note added:} Following submission of this work, the closely related preprint \cite{Greydanus:2019aa} appeared.

In this work, we will focus on integrable models, which have nontrivial symmetries, as a test case for symmetry discovery. The organization of the remainder of this paper is as follows. In the next section, we introduce some concepts from classical integrable systems that we will use. \Cref{sec:deep} introduces our new architecture and learning algorithm, while \cref{sec:exp} contains experiments on three integrable models. Finally, \cref{sec:disc} provides a discussion of the results. Supplementary details are contained in the appendices.

\section{Classical integrable systems}
\label{sec:2}

\subsection{Hamiltonian dynamics and canonical transformations}

Classical mechanics is the realm of Hamiltonian dynamics \cite{Saletan:1998aa}. In the simplest case that we address here, motion occurs in a phase space $\mathbb{R}^{2n}$ of positions $q\in \mathbb{R}^{n}$ and momenta $p\in \mathbb{R}^{n}$. The dynamics is governed by Hamilton's equations, which are derived from a Hamiltonian function $H:\mathbb{R}^{2n}\to\mathbb{R}$. For $x = (q,p)$ these read:
\begin{align}
  \label{eq:eom}
  \dot{x} = \Omega \nabla_x H\, ,\quad
  \Omega =
  \begin{pmatrix}
    0 & \id_n\\
    -\id_n & 0
  \end{pmatrix}
  \,.
\end{align}
For example, the harmonic oscillator with unit frequency is described
by $H = ( p^2 + q^2 )/2$, and the equations of motion $\dot q = p$, $\dot p = -q$ describe circular orbits in the phase plane.

\begin{figure}[hbtp]
  \centering
  \includegraphics[width=\columnwidth]{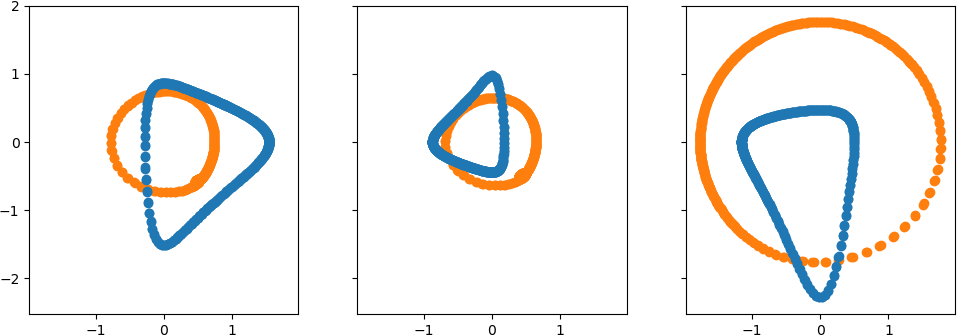}
  \caption{Projections of the original (blue) and transformed (orange) trajectories of
    the Kepler Hamiltonian (\cref{eq:HKepler} with $k=-1$) onto the three
    $q_j-p_j$ phase planes. Note that the sum of enclosed areas is the same for the original and transformed trajectories.}
  \label{fig:1d_example}
\end{figure}

The skew-symmetric matrix $\Omega$ is called a symplectic form, and allows
one to define symplectic (or canonical) transformations of phase
space, as those maps $f$ whose Jacobian matrix $J_f$ is an element of
the linear symplectic group $\Sp_{2n}(\mathbb{R})$ at each point of its domain:
\begin{align}\label{eq:symp_cond}
J_f^T\Omega J_f = \Omega \,.
\end{align}
Since $\Det(J_f) = +1$, volume is conserved. \Cref{eq:symp_cond} is however much more restrictive: the sum of (signed) areas in each $q_j-p_j$ plane is preserved (\cref{fig:1d_example}).

Given $u,v$ scalar valued functions on phase space, their Poisson
bracket is defined as $ \{ u, v\} \equiv (\nabla_x u)^\top \Omega \nabla_x v$.
and is a symplectic invariant. With this notation, Hamilton's
equations are $\dot{x} = \{ x, H \}$, and time evolution is itself a
symplectic transformation \cite{Arnold:2013aa}.

\subsection{Integrable models}

A conserved quantity of a dynamical system is constant in time, and
thus Poisson--commutes with the Hamiltonian, and constitutes a symmetry
of the problem. For example, in the celebrated Kepler problem
describing the motion of two planets attracted by a gravitational
force which depends only on the distance between the planets, $H$
commutes with the angular momentum, which generates rotations of the
relative coordinate, and is a conserved quantity of the dynamics
\cite{goldstein}.

\emph{Integrable} systems are those which have a number of mutually
commuting and independent integrals of the motion that equals $n$,
half the phase space dimension.  The Liouville--Arnold theorem states
that (compact) motion is confined to torii parametrized by angles
$\varphi_1,\dots,\varphi_n$ and there exists a symplectic
transformation $\mathcal{T}^{-1}$ from the original coordinates $q,p$
to new coordinates $\varphi,I$, where $I$ are called actions and are
the conserved quantities of the problems \cite{Arnold:2013aa}.  In the
action angle coordinates, \cref{eq:eom} therefore reads:
\begin{align}
  \label{eq:eom_action_angle}
  \dot{\varphi} = \partial_IK = \text{const.}\,,\quad
  \dot{I} = -\partial_\varphi K = 0
  \,,
\end{align}
where the transformed Hamiltonian
\begin{align}
  \label{eq:K}
  K = H \circ \mathcal{T}
  \,,
\end{align}
is independent of the angles.  Finding explicit action-angle transformations is
a challenging task and while a lot of progress has been made
constructing integrable systems from algebraic or geometric
principles \cite{babelon2003}, there is no general algorithm to
construct higher integrals of the motion given an integrable
Hamiltonian. Learning such a transformation is the goal of this work.

We will work with the Cartesian coordinates $(\hat{q}_i =
\sqrt{2I_i}\cos(\varphi_i), \hat{p}_i = \sqrt{2I_i}\sin(\varphi_i))$ and denote
by $T$ the symplectic map
\begin{align}
  \label{eq:T}
  T : (\hat{q}, \hat{p}) \mapsto (q, p) \,.
\end{align}
For example, action-angle variables for the harmonic oscillator are
the symplectic polar coordinates $(\arctan(p/q), (p^2+q^2)/2)$, so that $T$ is the identity, and the
only conserved quantity is the energy.
In general $T$ will be such that complex trajectories in the $(q,p)$
phase space get mapped to circles in the $(\hat{q}_i, \hat{p}_i)$ planes
where $\varphi_i$ is the angular coordinate and $I_i$ is half the squared
radius of the circle.

In this work we will learn neural parametrizations of $T$ for three paradigmatic
integrable models: (1) the Kepler model of two planets interacting with
gravitational force \cite{goldstein}; (2) the Neumann model of $n$
oscillators with positions in $\mathbb{R}^n$ constrained to the
$(n-1)$--dimensional sphere \cite{babelon2003}; (3) the Calogero-Moser (CM)
model of a chain of $n$ particles with inverse square interaction potential \cite{MOSER:1976aa}.  The Hamiltonians and their conserved quantities are described in \cref{sec:Ham}.

% Even though many models of interest in physics are not integrable,
% studying integrable models is important since it allows one to make
% precise comparisons between theory and experiments \cite{??} and to
% study non-integrable models as perturbation of integrable ones
% \cite{??}.
%
% TODO Additionally, many models display approximate conservation laws e.g. FPU. Generalized Gibbs

% 3 cols = 1+1/2 + 1+1/2
\section{Deep symplectic flows} \label{sec:deep}

We have reformulated the task of learning symmetries in integrable
models as that of learning the map $T$ which transforms a circular trajectory $(\hat{q}_i(t),\hat{p}_i(t))$ to the complex trajectory of the original model $(q_i(t),p_i(t))$. We now describe how to parametrize and learn such a transformation.

\subsection{Parametrization}
\label{sec:3.1}

Here we will adapt recent results on normalizing flows to provide
symplectic versions of popular invertible layers such as additive
coupling \cite{NICE}, batch normalization \cite{realNVP} and invertible linear
transformations \cite{glow}.  Our parametrization of $T$ is given by
stacking $m$ blocks of these three layers. %as depicted in \cref{fig:arch} and
We now describe each layer.

% \begin{figure}[h]
%   \centering
%   \input{figures/dsf.tex}
%   \caption{Network architecture. Do we need this?}
%   \label{fig:arch}
% \end{figure}

%% Symplecticity of the mapping can be seen as an inductive bias on these
%% network architectures.

\subsubsection{Symplectic additive coupling}

The additive coupling layer introduced in \cite{NICE}
partitions the inputs $z = (z_A, z_B)$
and outputs $x = (x_A,x_B)$ with $x_A = z_A, x_B = z_B + \NN(x_A)$,
where the shift function $\NN$ is an arbitrary neural network.  If we
now identify $A,B$ subsystems as $q,p$ respectively, we have the following
layer $L : (q,p) \mapsto (Q,P)$:
\begin{align}
  (Q, P) = (q, p + \NN(q))\, ,\quad
  (q, p) = (Q, P - \NN(Q))\,.
\end{align}
Symplecticity of the transformation imposes further irrotationality:
\begin{align}
  \label{eq:irrot}
  \partial_i \NN_j = \partial_j \NN_i\, .
\end{align}
This constraint may be handled by setting $\NN(q) = \nabla
F(q)$, where $F:\mathbf{R}^n\to \mathbf{R}$ is parametrized by a neural network. This gives the traditional leapfrog update used in HMC.

While conceptually simple, this approach is computationally
expensive, requiring $O(n^2)$ time in the backward pass of the
network.  A cheaper approach is to use a multilayer perceptron with three layers and constrained weight matrices $W^a$ $a=1,2,3$. In \cref{sec:imlp} we show that \cref{eq:irrot} is satisfied if
\begin{align}\label{eq:weight_cond}
  W^3 = W^{1\top}\, ,\quad
  W^{2} = \diag(w^2_1, \dots, w^2_{n_2})\, .
\end{align}
We call this architecture irrotational MLP.  The analysis can be
generalized, but we took this simple architecture for most of our
experiments. Geometrically, $W^1$ embeds $q$ into a higher dimensional
space and $W^{1\top}$ maps the embedding back to the original space
after it has been scaled by $W^2$, whose sign controls whether the map
is orientation preserving.

\subsubsection{Symplectic Linear}

The additive coupling leaves $q$ unchanged, and we introduce the
symplectic linear layer to mix $p,q$ so that deeper additive couplings act
on all phase space coordinates.
To parametrize a symplectic matrix $S\in \Sp_{2n}(\mathbb{R})$,
we use the pre--Iwasawa decomposition \cite{deGosson}:
\begin{align}
  S =
  NAK=
  \begin{pmatrix}
    \id & 0 \\
    M & \id
  \end{pmatrix}
  \begin{pmatrix}
    L^\top & 0 \\
    0 & L^{-1}
  \end{pmatrix}
  \begin{pmatrix}
    X & -Y \\
    Y & X
  \end{pmatrix}
  \, ,
\end{align}
with
\begin{align}
  &M = M^\top
  \,,\quad
  &X^\top Y = Y^\top X \, ,\quad
  X^\top X + Y^\top Y = \id\,.
\end{align}
To parametrize $K$, we note that it is the realification of the
unitary $X + iY$ and can be written as a product of
Householder reflections, parametrized by a vector $v$:
\begin{align}
  R_v = \id - 2 \frac{v v^\dagger}{||v||^2}\in {\rm U}_n\,,
\end{align}
and a diagonal matrix of phases
\begin{align}
U = \diag(e^{i\phi_i})\,.
\end{align}
We refer to \cite{Cabrera2010,Tomczak16} for background on Householder
reflections.

Note that the complexity of applying both $R_v$ and $U$ to a vector
$(q,p)$ is $O(n)$. To keep the complexity of the whole layer $O(n)$ we
take $r = O(1)$ Householder reflections and further take
$L=\diag(L_i)$ and $M = \diag(M_i)$.

% The full operation is
% \begin{align}
%   S &=
%   \begin{pmatrix}
%     \diag(L_i) & 0 \\
%     \diag(M_iL_i) & \diag(1/L_i)
%   \end{pmatrix}
%   R_{v_1}\cdots R_{v_r}
%   U
%   \\
%   S^{-1} &=
%   U^{-1}
%   R_{v_r}\cdots R_{v_1}
%   \begin{pmatrix}
%     \diag(1/L_i) & 0 \\
%     -\diag(L_iM_i) & \diag(L_i)
%   \end{pmatrix}
%   \, .
% \end{align}

\subsubsection{Zero center}

The zero center layer is defined by the transformation:
\begin{align}
  \begin{cases}
    Q = q - \mu^q  + \alpha\\
    P = p - \mu^p  + \beta
  \end{cases}\, ,
\end{align}
where $\mu$ is the mean given during training as the batch mean, and
during testing by a weighted moving average accumulated during
training, as in the batch normalization layer -- see \cite{realNVP}
for its usage in normalizing flows. $\alpha$, $\beta$ are learnable
offsets.  The zero center layer normalizes its input and allows one to
study deeper architectures, and is a restricted version of batch
normalization compatible with symplectic invariance.  (The full
version of batch normalization indeed scales by the variance of each
feature and does not preserve areas.)

\subsection{Learning algorithm}
\label{sec:3.2}

% Placing it here so that it appears on page 4...

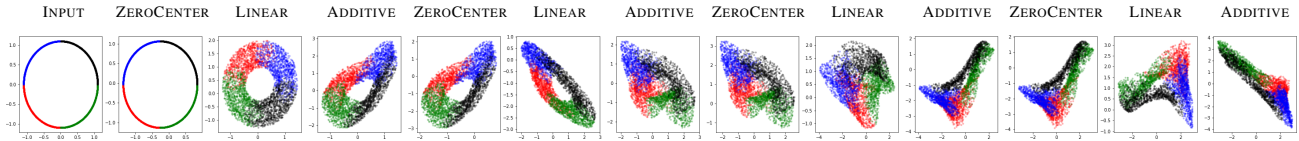
\begin{figure*}[hbtp]
  \centering

  \begin{tikzpicture}

    \node[] at (0,0) {
      \includegraphics[width=\textwidth]{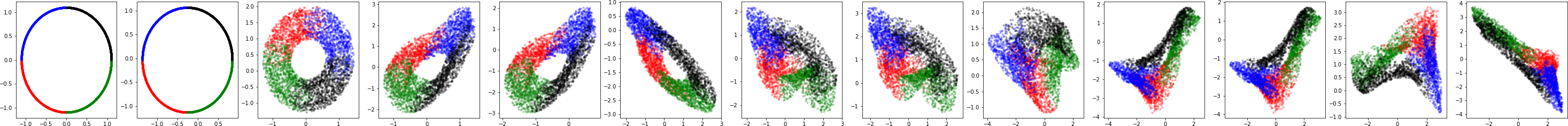}};

    \begin{scope}[xshift=-6.5cm,yshift=1cm]
      \newcommand\dx{1.32}
      \node[] at (-\dx,0) {\textsc{\tiny Input}};
      \foreach \i in {0,3,6,9}
               {
                 % zero center
                 \node[] at (\dx * \i,0) {\textsc{\tiny ZeroCenter}};

                 % linear
                 \node[] at (\dx * \i + \dx,0) {\textsc{\tiny Linear}};

                 % additive
                 \node[] at (\dx * \i + 2*\dx,0) {\textsc{\tiny Additive}};
               }
    \end{scope}

  \end{tikzpicture}

  \caption{Visualization of the transformations done by each layer of
    $T$ along a given $q_j-p_j$ phase plane for the Neumann model.
    The input points (left) belong to a cycle of the Liouville--Arnold
    torus. Colors refer to which quadrant of the input plane a point
    comes from.}
  \label{fig:viz_neumann}
\end{figure*}

% \begin{figure*}[hbtp]
%   \centering
%   \begin{subfigure}{0.3\textwidth}
%     \centering
%     \includegraphics[width=\textwidth]{figures/kepler_t2c}
%     \caption{Kepler ($k=-1$)}
%     \label{fig:Kepler}
%   \end{subfigure}
%   \hfill
%   \begin{subfigure}{0.3\textwidth}
%     \centering
%     \includegraphics[width=\textwidth]{figures/neumann_t2c}
%     \caption{Neumann($a_1=.1,a_2=.2,a_3=.3$)}
%     \label{fig:Neumann}
%   \end{subfigure}
%   \hfill
%   \begin{subfigure}{0.3\textwidth}
%     \centering
%     \includegraphics[width=\textwidth]{figures/CM_t2c}
%     \caption{Calogero-Moser ($\omega^2 = g^2 = 1$)}
%     \label{fig:CM}
%   \end{subfigure}
%   \caption{Pull back of trajectories (thin blue line) under the
%     action-angle map $T^{-1}$ learned using the traj2circles
%     algorithm. Each model has phase space of dimension $2n=6$ and here a single
%     $q-p$ phase plane is selected for illustration. TODO Looks like we could fit this in one column with a longer caption?}
%   \label{fig:2}
% \end{figure*}

According to the discussion in \cref{sec:2}, the map $T^{-1}$ is
determined by requiring that the original trajectory $(q_i(t),p_i(t))$ is
mapped to circles $(\hat{q}_i(t),\hat{p}_i(t))$. If the trajectory is
sampled at $\tau$ time steps $t_k$, such $T$ minimizes the
following loss, which encourages the distance from the origin of
neighbouring points to be the same:
\begin{align}
  \label{eq:loss}
  \ell =
  \frac{1}{n \tau}
  \sum_{k=1}^{\tau}
  || r_{k} - r_{k+1} ||^2\,,\quad
  r_{k} =
  \hat{q}(t_k)^2 +
  \hat{p}(t_k)^2 \,.
\end{align}
A non-invertible or non-volume preserving network
could minimize \cref{eq:loss} trivially by collapsing the trajectories
to zero or very small volumes in the transformed phase space: the symplecticity of the transformation is essential.

We therefore consider a learning algorithm that
takes as input a batch of trajectories and minimizes the loss above
averaged over the batch.  Practically, we compute the trajectories by
solving the original equations of motions using a Runge-Kutta solver,
and we perform stochastic gradient descent parameter updates using the
Adam optimizer. We shuffle randomly the trajectories at every epoch,
which ensures that we compare distant points in \cref{eq:loss}, so that
all deviations from a circular shape are penalized in the same way.

\section{Experiments} \label{sec:exp}
\label{sec:4}

\begin{figure}[hbtp]
  \centering
  \includegraphics[width=0.45\columnwidth]{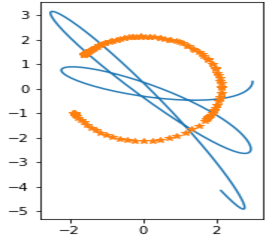}\includegraphics[width=0.45\columnwidth]{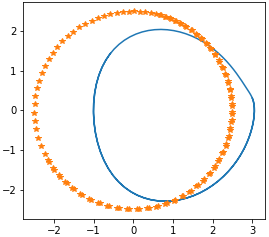}
  \caption{Pull back of trajectories (thin blue line) under the map
    $T^{-1}$ learned. Each model has phase space of dimension $2n=6$
    and here a single $q-p$ phase plane is selected for
    illustration.}%
  \label{fig:2}
\end{figure}

We now present the results of experiments on the three models defined
in \cref{sec:Ham}. The code used is available at
\textbf{\texttt{\url{https://github.com/rbondesan/CanonicalFlows/}}},
and we refer to \cref{sec:details} for details of network
architecture and other training hyperparameters.

We first discuss the representation capacity of deep symplectic
flows. \Cref{fig:2} shows the pull-back of the trajectories under a
network $T^{-1}$ composed of $m=4$ blocks defined in \cref{sec:3.1}.
The parameters in $T^{-1}$ are learned by running the algorithm of
\cref{sec:3.2} to convergence and feeding it with a single trajectory
sampled at $\tau=128$ time steps. This shows that our model and
algorithm can learn the action--angle map for both periodic (Kepler
and Calogero-Moser) and quasi-periodic (Neumann) models.

\begin{figure*}[hbtp]
  \centering
  \includegraphics[width=\textwidth]{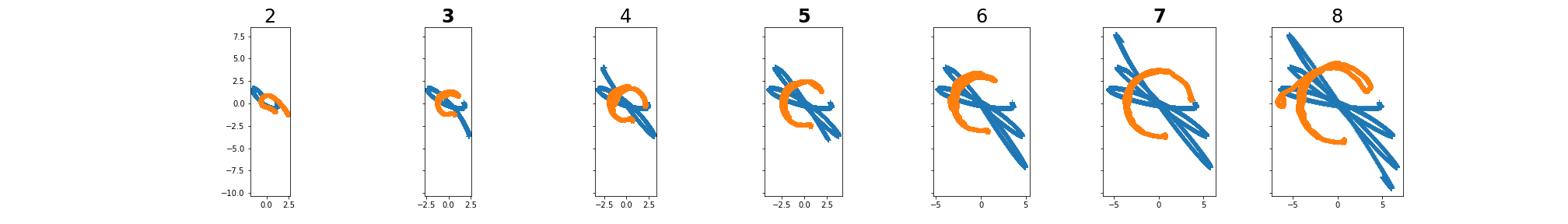}
  \caption{Pull back of trajectories (thin blue line) for the Neumann
    model at radius $r$, indicated by the figures titles.  Bold text
    denotes radii of training trajectories. A single $q-p$ phase plane
    is selected for illustration.}
  \label{fig:neumann-2-seen-unseen-traj}
\end{figure*}

We next investigate the generalization of our learning algorithm.  By
this we mean how well the learned symplectic flow can map unseen
trajectories to circles. We present here results for the Neumann model
with $n=3$ oscillators. We consider a batch of trajectories, one for
each radius $r=2,3,\dots,8$ of the sphere on which the positions of
the oscillators move.  \Cref{tab:neumann} reports the loss of
\cref{eq:loss} evaluated over these trajectories for a symplectic flow
$T$ learned by considering only the subset $r=3,5,7$ as training
data. While the points in the training set have the smallest values
compared to neighbouring radii, the fact that the loss is of the same
order across this range of trajectories, indicates that the model
generalizes beyond the training points. To further substantiate this
claim, we show in \cref{fig:neumann-2-seen-unseen-traj} the pull-back
of the trajectories under $T^{-1}$ for both the trajectories seen and
unseen by the training algorithm.

\begin{table}[t]
  \caption{Test loss $\ell'=\ell\times 10^{5}$ for trajectories in
    the Neumann model at radius $r$. Bold text denotes radii of training trajectories.}
\label{tab:neumann}
\vskip 0.15in
\begin{center}
\begin{small}
\begin{sc}
\begin{tabular}{rccccccccr}
\toprule
$r$ & 2 & \textbf{3} & 4 & \textbf{5} & 6 & \textbf{7} & 8 \\
\midrule
$\ell'$ &
%20.8 &
6.5 &
\textbf{3.7} &
23.2 &
\textbf{12.4} &
117.5  &
\textbf{23.4}  &
141.6 \\
\bottomrule
\end{tabular}
\end{sc}
\end{small}
\end{center}
\vskip -0.1in
\end{table}

The map $T$ thus learned can be used as a generative process for the
physical system trajectories, as illustrated in
\cref{fig:viz_neumann}.  Interestingly, this allows one to visualize
the points in phase space that correspond to a given Liouville--Arnold
torus. Further, by varying the circles radii one can interpret the effect
of the learned symmetries on the system under consideration.

% 1/2 col
\section{Discussion}\label{sec:disc}

The learning algorithm discussed so far relies on being able to solve
the equations of motion. This was done here by numerical integration,
and adds questions of convergence and stability of the ODE solver on
top of those of the learning algorithm. We remark that there two
possible ways to improve this.  The first is to exploit
integrability of the models to solve the motion analytically, which is
however not a trivial matter and typically requires uncovering a Lax
pair formulation of the problem \cite{babelon2003}.  The second is to
use a different learning algorithm. For example, one could minimize a
loss that encourages the transformed Hamiltonian to be
angle-independent (recall \cref{eq:eom_action_angle}), or one could
minimize the Kullback--Leibler (KL) divergence between the canonical
density associated to the transformed Hamiltonian and that of a base
distribution. Both alternatives are analyzed in \cref{sec:alt}.

While we have concentrated here on integrable models, we expect our
methods to be applicable to find almost conserved quantities in models
close to integrable ones, such as the celebrated
Fermi--Pasta--Ulam--Tsingou chains \cite{Gallavotti:2007aa}.  We thus
expect the deep learning approach to classical mechanics presented
here to be of practical relevance for solving physical problems, for
example by finding integration schemes with smaller discretization
errors.

% \section*{Acknowledgements}\label{sec:ack}

\bibliographystyle{icml2019}

\appendix

\section{Symplecticity of irrotational MLP}\label{sec:imlp}

We demonstrate that the stated form of the weights in
\cref{eq:weight_cond} for a three layer MLP satisfies the
irrotationality condition of \cref{eq:irrot}. The Jacobian of the MLP
is
\begin{align}
  \frac{\partial \text{NN}(x) }{\partial x }
  =
  W^3 D^{2} W^2 D^1 W^1\, ,
\end{align}
with $D^a = \diag(\lambda^a)$, $\lambda^a = (\sigma^{a})'( W^a x^a +
b^a)$, where $W^a, b^a, \sigma^a, x^{a+1}$ are resp.~$n_{a}\times n_{a-1}$
weights, bias, activation functions and activations of the layer
$a$. The constraint of \cref{eq:irrot} then takes the form
\begin{align}\label{eq:sym_cond}
  W^3 D^{2} W^2 D^1 W^1
  =
  W^{1 \top} D^{1} W^{2\top} D^2 W^{3\top}\, .
\end{align}
We further take $n_2 = n_3$, so that $D^{2} W^2 D^1$ is square.  The
symmetry condition \cref{eq:sym_cond} means that $W^{1\top}=W^3$ and $D^{2} W^2 D^1$ is symmetric. For generic $\lambda^1, \lambda^2$, the only solution is $W^2_{ij}=0$ for $i\neq j$. This verifies \cref{eq:weight_cond}.

\section{Hamiltonians and conserved quantities}\label{sec:Ham}

We provide details of the three integrable models studied.

\subsection{Kepler}

The Hamiltonian of the Kepler model is
\begin{equation}
\label{eq:HKepler}
H_{\text{K}} = \tfrac{1}{2}\sum p_i^2 + \frac{k}{r}\,,
\quad r = \sqrt{\sum q_i^2}.
\end{equation}
As well as the Hamiltonian $H_\text{K}$ and the three components of the angular momentum $\mathbf{L}=\mathbf{q}\times\mathbf{p}$, the Kepler model has an additional conserved quantity called the \emph{Laplace--Runge--Lenz vector} \cite{goldstein}
\begin{equation}
  \mathbf{A} = \mathbf{p}\times\mathbf{L} + k \frac{\mathbf{q}}{r}.
\end{equation}
This gives a total of seven conserved quantities. However, a one dimensional trajectory in a six dimensional phase space can have at most five conserved quantities that specify the trajectory. Thus there must be two relations between these quantities. They are
\begin{align}
\mathbf{A}\cdot \mathbf{L}&=0\nonumber\\
\mathbf{A}^2 &= k^2 + 2H_\text{K} \mathbf{L}^2
\end{align}
The existence of five independent conserved quantities, two more than the three required for integrability, mean that the trajectories form closed curves rather than filling the Liouville--Arnold torii. This situation is called \emph{superintegrability}.

\subsection{Neumann}

The Neumann Hamiltonian is
\begin{equation}
  H_{\text{N}} =
  \tfrac{1}{4} \sum_{i,j} J_{ij}^2 + \tfrac{1}{2} \sum a_i q_i^2\,,
  \quad
  J_{ij} = q_i p_j - q_j p_i.
\end{equation}
An independent set of constants of motion are \cite{babelon2003}
\begin{equation}
I_i = q_i^2 + \sum_{i\neq j} \frac{J_{ij}^2}{a_i-a_j}
\end{equation}
for $k_\alpha$ all different. The property $\{I_i,I_j\}=0$ for $i\neq j$ is easily checked using $\{q_i,p_j\}=\delta_{ij}$. The Hamiltonian can be expressed as the linear combination
\begin{equation}
H_\text{N} = \frac{1}{2}\sum_i a_i I_i.
\end{equation}
Additionally,
\begin{equation}
\sum_i q_i^2 = \sum_i I_i,
\end{equation}
showing that motion is confined to the sphere $S^{n-1}$.

\subsection{Calogero--Moser}

The Calogero--Moser model is given by
\begin{equation}
  \label{eq:HCM}
  H_{\text{CM}} =
  \tfrac{1}{2}\sum (p_i^2 + \omega^2 q_i^2) +
  \sum_{j < k} \frac{g^2}{(q_j - q_k)^2}\,.
\end{equation}
The discussion of integrals of motion is facilitated by presenting the equations of motion in matrix form \cite{Perelomov:1990aa}. Introducing the matrices
%
% https://arxiv.org/pdf/1103.6231.pdf
%
\begin{align}
  L^\pm &= L \pm i\omega Q\nonumber\\
  Q_{ij} & = q_i\delta_{ij}\nonumber\\
  L_{ij} &= p_i\delta_{ij}+(1-\delta_{ij})\frac{ig}{q_i-q_j}\nonumber\\
  M_{ij} &=  g\left[\delta_{ij}\sum_{k\neq i}\frac{1}{(q_i-q_k)^2} - (1-\delta_{ij})\frac{1}{(q_i-q_j)^2}\right],
\end{align}
the equations of motion are equivalent to
\begin{align}
  \dot Q + i[M,Q] = L\nonumber\\
  \dot L + i[M,L] = -\omega^2 Q.
\end{align}
From this point it can be shown that the quantities
\begin{equation}
  I_k = \mathop\mathrm{tr}\left[(L_+L_-)^k\right]
\end{equation}
are all conserved.

\section{Details of experiments}
\label{sec:details}

We present here details of the experiments of \cref{sec:4}.

In \cref{fig:2,fig:viz_neumann} we used frequencies
$a_1=.1,a_2=.2,a_3=.3$ for the Neumann model and couplings $\omega^2 =
g^2 = 1$ for the Calogero-Moser model.  The network architecture used
is composed of $m=4$ blocks of zero-center, linear symplectic and
symplectic additive coupling layers of \cref{sec:3.1}, applied in this
order.  The linear symplectic layer has $r=2$ Householder reflections
whose reflection vectors are initialized to be identical, so that
their product is the identity. The irrotational MLP used in the
additive coupling layer (see \cref{sec:imlp}) has hidden dimension
$512$ and $\tanh$ activation functions.

The data reported in \cref{tab:neumann} and
\cref{fig:neumann-2-seen-unseen-traj} was produced by sampling at
$\tau=4096$ points solutions of the equations of motions between times
$0$ and $10$. Training was done with the Adam optimizer using a
mini-batch size of $128$. We trained for $7395$ epochs using a
piece-wise learn rate schedule with iteration step boundaries $[20000,
  50000, 100000]$ and values $[10^{-3}, 10^{-4}, 10^{-5}, 10^{-6}]$.

\section{Alternative algorithms}
\label{sec:alt}

\subsection{\texttt{dKdPhi}}
\label{sec:dKdPhi}

The algorithm of \cref{sec:3.2} used numerical solutions of the
equations of motion for a batch of initial conditions. This can be
expensive for large $n$ and also introduces numerical errors and
questions of convergence of the ODE solver.  We here present an
alternative procedure.  Recall that the transformed Hamiltonian $K$ of
\cref{eq:K} satisfies $\nabla_\varphi K = 0$. We therefore could find
$T$ by minimizing the expectation of the norm of $\nabla_\varphi
K$. However, the natural expectation is with respect to the canonical
density of the model, which is $\propto \exp(-K(I))$ and itself
unknown.  We propose to replace the canonical density with the
exponential density
\begin{align}
  \label{eq:exp_density}
  \rho(\varphi, I) = (2\pi)^{-n} \exp -\sum_i I_i \,,
\end{align}
where the numerical prefactor comes from the uniform density of the
angles.  This choice is motivated by the fact that integrable models
are expected to be described by the generalized Gibbs ensemble when
$n$ is large \cite{Jaynes:1957aa}, and which corresponds to an exponential
distribution for the actions.  Therefore, we propose the following
alternative loss:
\begin{align}
  \label{eq:alt_loss}
  \ell &=
  \mathbb{E}_{(\varphi,I)\sim \rho}
  \{ || \nabla_\varphi K ||^2 \}
  \,,
\end{align}
where we estimate the expectation over minibatches.

\begin{figure}[hbtp]
  \centering
  \input{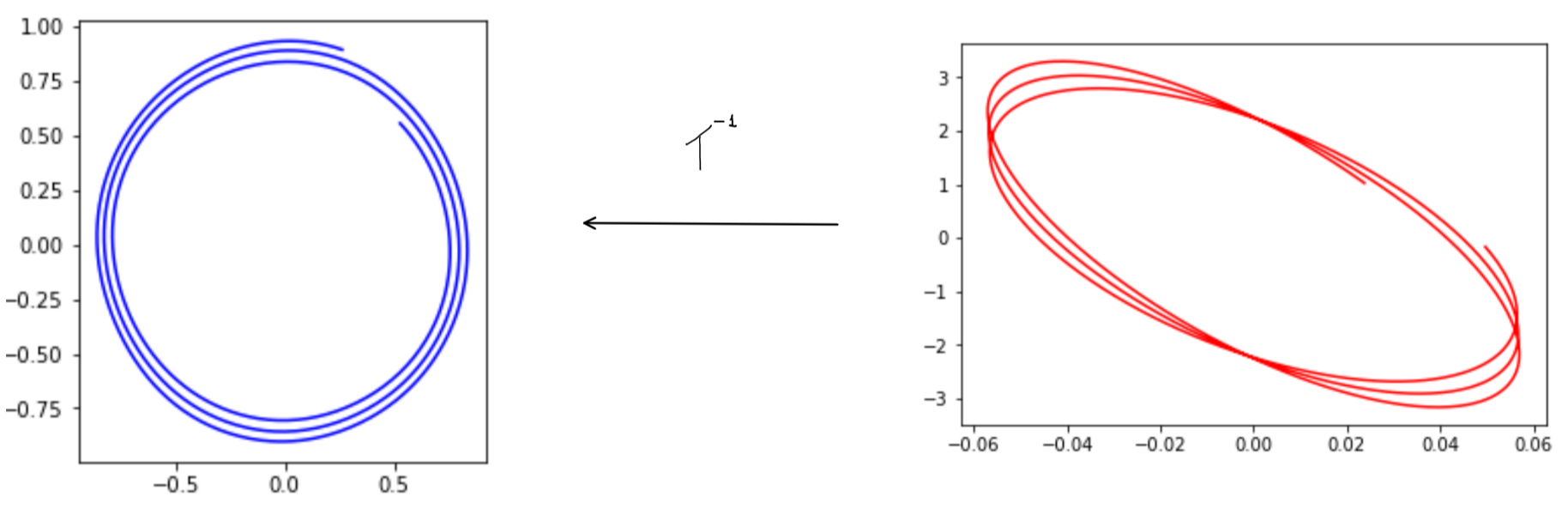}
  \caption{Pull back of Neumann trajectories (right) under the map
    $T^{-1}$ learned by minimizing \cref{eq:alt_loss} for a single $q-p$
    phase plane.}
  \label{fig:dKdPhi_neumann}
\end{figure}

For illustration, we present some results using this algorithm for the
Neumann model. For simplicity, we trained the model by sampling
uniformly in the angles but fixing a single value of the actions,
effectively replacing the exponential measure in $\rho$ with a Dirac
measure, which physically amounts to restricting sampling to a single
Liouville--Arnold torus. The model parameters and network architecture
are similar to those of \cref{sec:details}.  To test the algorithm, we
solve the Neumann equations of motion for an initial condition
corresponding to the value of the actions used for training, and check
that the inverse map is able to trivialize the trajectory by mapping
it to a circle. We stress that differently from \cref{sec:4}, here the
trajectories are not inputs to the training algorithm, which is
fully unsupervised. The result is presented in
\cref{fig:dKdPhi_neumann}, and validates the procedure.

While not relying on numerical solutions of the equations of motion is
appealing, we noticed that this algorithm takes a longer time to
converge than that of \cref{sec:3.1}.  This is partly due to the
presence of a derivative in the loss, which increases the
computational cost.

\subsection{Normalizing flows and KL minimization}

The algorithm of \cref{sec:dKdPhi} relies on the exponential ansatz
for the sampling measure, which lacking rigorous results about its
validity, can be not optimal in practice. Also, while one
learns directly the transformed Hamiltonian $K$, one does not know how
to sample efficiently from it.  Both issues can be resolved if we take
a learnable density
\begin{align}
  \rho(I,\varphi) = \rho_0(F^{-1}(I), \varphi) \left| \Det(\partial_{I} F^{-1}(I) ) \right|\,,
\end{align}
where $F$ is a normalizing flow \cite{realNVP}. The base distribution
$\rho_0$ can be taken to be the exponential distribution.

In this formulation, a natural loss to consider is the following KL
divergence:
\begin{align}
  \label{eq:loss_KL}
  \ell &=
  \mathbb{E}_{(u,\varphi)\sim \rho_0}\left( \text{KL}(\rho_0, \pi\circ T \circ F) \right) \\
  &\sim
  \frac{1}{N}
  \sum_{i=1}^N H(T(F(u_i),\varphi_i)) + \text{const}
  \, ,
\end{align}
where $\pi \propto \exp(-H)$ is the canonical density associated to
the Hamiltonian $H$.  We leave the numerical study of the minimization
of this loss for future work.

\end{document}